\documentclass[aps,prl,twocolumn,superscriptaddress,amsfont,amssymb,amsmath,nofootinbib,showpacs,balancelastpage]{revtex4-1}
\usepackage{graphicx,longtable,natbib,mathrsfs,color}

\newcommand{\om}{{\Omega_m}}
\newcommand{\ob}{{\Omega_b}}

\newcommand{\odm}{{\Omega_\mathrm{DM}}}

\newcommand{\ho}{{H_0}}

\newcommand{\lcdm}{$\Lambda$CDM}

\newcommand{\fnl}{{f_\mathrm{NL}}}
\newcommand{\fov}{\mathrm{FoV}}

\begin{document}

\title{Cosmology on Ultralarge Scales with Intensity Mapping of the Neutral Hydrogen 21 cm Emission: Limits on Primordial Non-Gaussianity}

\author{Stefano Camera}
\email{stefano.camera@ist.utl.pt}
\author{M\'ario G. Santos}
\email{mgrsantos@ist.utl.pt}
\affiliation{CENTRA, Instituto Superior T\'ecnico, Universidade T\'ecnica de Lisboa, Av. Rovisco Pais 1, 1049-001 Lisboa, Portugal}
\affiliation{Department of Physics, University of the Western Cape, Bellville 7535, South Africa}
\author{Pedro G. Ferreira}
\email{p.ferreira1@physics.ox.ac.uk}
\affiliation{Astrophysics, University of Oxford, DWB, Keble Road, Oxford OX1 3RH, UK}
\author{Lu\'is Ferramacho}
\affiliation{CENTRA, Instituto Superior T\'ecnico, Universidade T\'ecnica de Lisboa, Av. Rovisco Pais 1, 1049-001 Lisboa, Portugal}
\affiliation{Department of Physics, University of the Western Cape, Bellville 7535, South Africa}

\date{Received \today; published -- 00, 0000}

\begin{abstract}
The large-scale structure of the Universe supplies crucial information about the physical processes at play at early times. Unresolved maps of the intensity of 21 cm emission from neutral hydrogen HI at redshifts $z\simeq1-5$ are the best hope of accessing the ultralarge-scale information, directly related to the early Universe. A purpose-built HI intensity experiment may be used to detect the large scale effects of primordial non-Gaussianity, placing stringent bounds on different models of inflation.  We argue that it may be possible to place tight constraints on the non-Gaussianity parameter $\fnl$, with an error close to $\sigma_\fnl\sim 1$. 
\end{abstract}


\maketitle

The statistical properties of large-scale structures are a rich source of information about the physics of the early Universe, its subsequent evolution and its current state (e.g., Ref.~\citep{Ade:2013ktc}). Constraints on the ultralarge-scale properties of density perturbations at redshifts $z\simeq1-5$ can improve our understanding of the primordial Universe; these scales are, in principle, uncontaminated by the nonlinear growth of structure or the poorly understood effects of baryons. Yet, such scales are also extremely difficult to access with conventional redshift or peculiar velocity surveys, while best attempts at using the cosmic microwave background are hampered by poor sampling, namely, cosmic variance.

An alternative approach for probing the density field of the large-scale structure has recently been advocated \citep{Battye:2004re,2008MNRAS.383..606W,Chang:2007xk,Peterson:2009ka}. It involves mapping out the combined emission of the $21$ cm, or HI, line from unresolved galaxies. In doing so, the large-scale structure is detected in three dimensions---a process which is usually referred to as ``intensity mapping.'' If one foregoes identifying individual galaxies, one can greatly speed up the observation and detection of the large-scale structure. Intensity mapping experiments are sensitive to structures at a redshift range that is observationally difficult to span for ground-based optical surveys \citep{Seo:2009fq}. Moreover, first attempts at mapping HI have been promising \citep{Chang:2010jp,Switzer:2013ewa}.

In this Letter we explore this method for constraining one particular aspect of current models of structure formation: the non-Gaussianity of primordial fluctuations, which may lead to scale dependent biasing and a distinctive signature on large scales. Furthermore, this approach can also be useful for probing general relativistic effects on these scales \citep{Bonvin:2011bg,Challinor:2011bk,Yoo:2013tc,Bruni:2011ta,Hall:2012wd}. Non-Gaussian initial fluctuations can arise in different models of inflation \citep{Maldacena:2002vr}. A particularly convenient (albeit not universal) way to parametrize non-Gaussianity consists of writing the gauge-invariant Bardeen potential $\Phi$---corresponding to the Newtonian potential in longitudinal gauge---as the sum of a Gaussian random field $\phi$ and a quadratic correction \citep{Komatsu:2001rj,Verde:1999ij}, $\Phi=\phi+\fnl\ast\left(\phi^2-\langle\phi^2\rangle\right)$, where $\ast$ denotes convolution 
between functions, and reduces to standard multiplication when $\fnl$ is a constant. Canonical single field inflation models predict $|\fnl|\sim\mathcal O(10^{-2})$ \citep{Maldacena:2002vr}, while evolution after inflation can generate an $\fnl\sim\mathcal O(1)$ \citep{Verde:1999ij,Liguori:2005rj,Smith:2006ud}. The method of excellence for constraining $\fnl$ has been to measure higher order correlation functions of the cosmic microwave background leading to $|\fnl|\lesssim 10^3$ with the MAXIMA data \citep{Santos:2001fg}, $|\fnl|\lesssim 10^2$ with the WMAP data \citep{Bennett:2012fp}, and now $|\fnl|\lesssim 10$ with the Planck data \citep{Ade:2013ydc}.

The non-Gaussian properties of initial fluctuations will also induce a scale and redshift dependence to a biased tracer $X$ of the underlying matter distribution \citep{Dalal:2007cu,Matarrese:2008nc}. The modification $\Delta b_X(z,k)$ to the Gaussian large-scale bias $b_X^G$ is such that $\Delta b_X(z,k)=3[b_X^G(z)-1]\om\ho^2\delta_c/[c^2k^2T(k)D_+(z)]\fnl$. Here, $\om=\ob+\odm$ is the total (baryons plus dark matter) matter fraction, $\ho$ is the Hubble constant, $\delta_c\simeq1.686$ is the critical collapse density contrast of matter, $T(k)$ is the matter transfer function versus the physical wave number $k$, and $D_+(z)$ is the linear growth factor of density perturbations. Attempts at detecting this effect with redshift surveys have led to some constraints on $\fnl$ \citep{Giannantonio:2013uqa}.

To assess how we might improve constraints on $\fnl$ using scale dependent bias, let us first restrict ourselves to a simple, cosmic variance limited survey. If we divide up the power spectrum in bins with constant logarithmic width, $\Delta k_\alpha=Ak_\alpha$, the number of modes per bin is given by $N_{\alpha}/V_\mathrm{eff}=Ak^3_\alpha/(2\pi^2)$, where $V_\mathrm{eff}$ is the effective volume of the survey and $\alpha$ spans the number of bins $N_\mathrm{bins}$. Then $\sigma^2_\fnl\simeq[b^G_XT(k_\alpha)D_+(z)c^2]^2/\{[6(b^G_X-1)\om\ho^2\delta_c]^2(\sum_\alpha N_\alpha k^{-4}_\alpha)\}$, where $T(k)$ is assumed constant on these large scales. Two obvious features immediately stand out: (i) better constraints on $\fnl$ will be obtained for larger $N_{\alpha}$ (that is for larger and deeper surveys), but also (ii) the further away the bias $b^G_X$ is from unity, the better. For fiducial values of the cosmological parameters and bias evolution, we find that $\sigma_\fnl\sim1$ may only be 
achievable if the survey depth is greater than $z\sim 3.5$. Intensity mapping surveys seem ideally suited for this goal (for proposals for doing so in the epoch of re-ionisation, see Refs.~\citep{Joudaki:2011sv,Hazra:2012qz,D'Aloisio:2013sda,Lidz:2013tra,Mao:2013yaa}).

When line-of-sight scattering and self-absorption phenomena are neglected, the HI line radiation discussed above can be related to the differential number counts of halo objects (e.g., Ref.~\citep{Challinor:2011bk}), from which we can estimate the HI bias $b_\mathrm{HI}$ \citep{Gong:2011ts}. The mean HI temperature from galaxies, assuming that the signal is seen in emission, is then $\overline{T}^g_b(z)\approx 566h\,[\ho/H(z)][\Omega_\mathrm{HI}(z)/0.003](1+z)^2\,\mu\mathrm K$, where $H(z)$ is the Hubble parameter, whose present-day value is $\ho=100h\,\mathrm{km\,s^{-1}\,Mpc^{-1}}$, $\Omega_\mathrm{HI}(z)\equiv \rho_\mathrm{HI}(z)/\rho_c$ is the comoving neutral hydrogen energy density in units of $\rho_c$, the critical density today. 

\begin{figure}
\centering
\vskip -.18in
\includegraphics[width=0.5\textwidth]{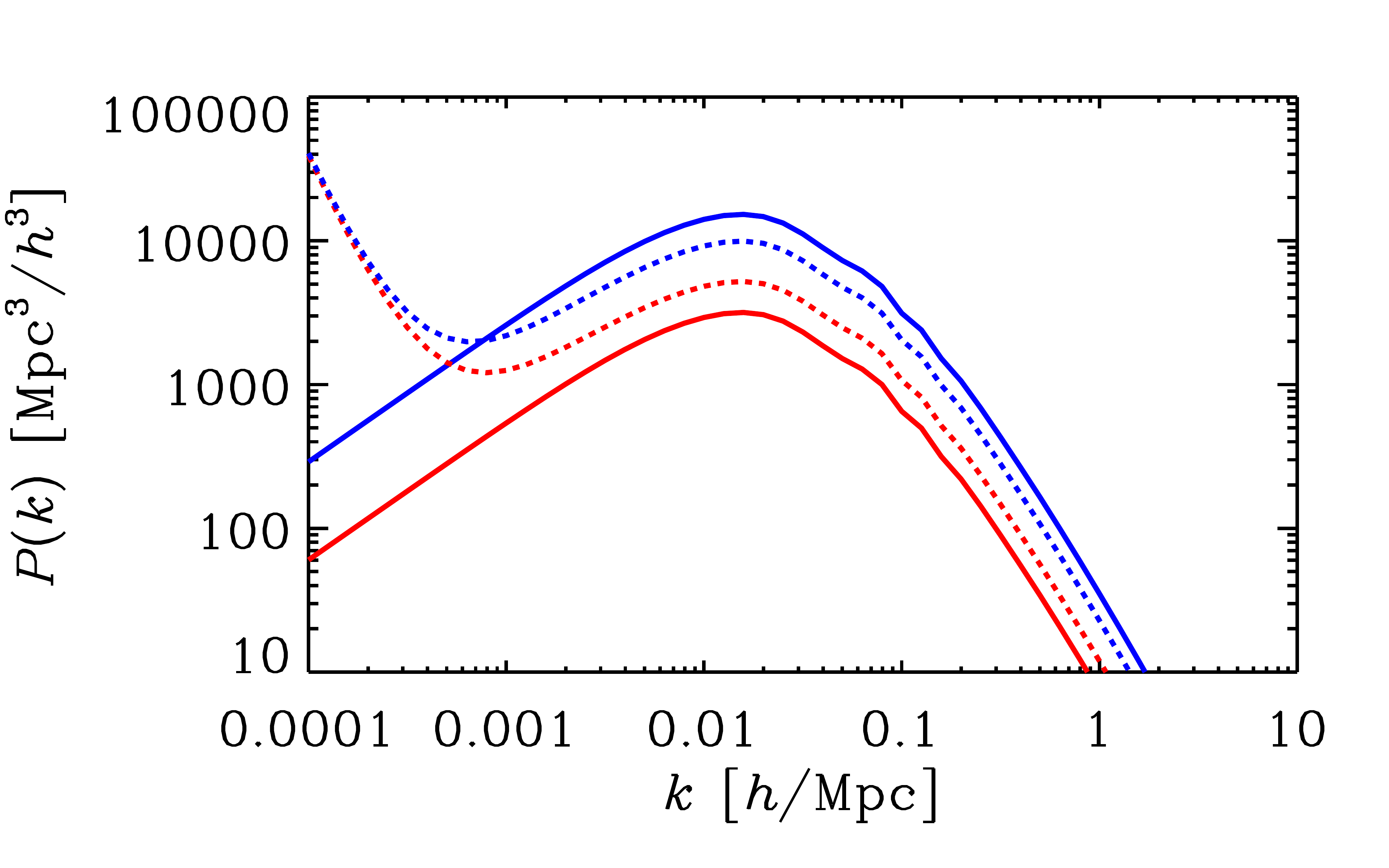}
\vskip -.2in
\caption{$P^\delta(k,z)$ (solid lines) and $b^2_\mathrm{HI}P^\delta(k,z)$ (dashed lines) for $|\fnl|=10$ at $z=0.4$ (top blue pair) and $z=2.5$ (bottom red pair).}\label{fig:P_k}
\vskip -.18in
\end{figure}

If we assume that, after reionization, the neutral hydrogen is mostly contained within galaxies, we calculate $\rho_\mathrm{HI}$ by integrating the Sheth and Tormen mass function \citep{Sheth:1999su}, assuming the HI mass to be proportional to the halo mass.  Setting the minimum and maximum mass in the integration by using a cutoff for the circular velocity \citep{Bagla:2009jy,Bagla:2009jy}, we can fix the constant of proportionality with the constraint on $\Omega_\mathrm{HI}(z)\times b_\mathrm{HI}(z)$ from Ref.~\citep{Switzer:2013ewa}. Finally we have that $b_\mathrm{HI}(z)$ is the appropriately weighted halo bias \citep{Sheth:1999su}. Thus the HI clustering power spectrum takes the form $P^\mathrm{HI}(k,z)=[\overline T^g_b(z)b_\mathrm{HI}(k,z)]^2P^\delta(k,z),$ with $P^\delta(k,z)$ the total matter power spectrum. The goal is then to target $P^\mathrm{HI}(k,z)$. In Fig.~\ref{fig:P_k}, we plot $P^\delta(k,z)$ and $b^2_\mathrm{HI}(z)P^\delta(k,z)$ for $|\fnl|=10$ (in synchronous gauge). We see that HI structures at low redshifts are underbiased with respect to dark matter, while at earlier times neutral hydrogen is highly biased. Moreover, the non-Gaussian effects we are looking for only come into play on extremely large scales.

To tackle the problem of forecasting $\sigma_\fnl$ more carefully, it is appropriate to work with the Fourier-Bessel transform on the sky and the  HI angular power spectrum $C^\mathrm{HI}_\ell(\nu_i,\nu_j)$, where $\nu_i$ is the frequency of shell $i$. To calculate this quantity, we use the \textsc{camb}$_-$sources code \citep{Challinor:2011bk}, and include the redshift space distortion but discard subdominant terms. Since, for an intensity mapping experiment, the frequency range can be (almost) arbitrarily small, the window function we adopt is thus a simple top-hat function. As a reference cosmology, we adopt a \lcdm\ flat universe with cosmological parameters $\om=0.28$, $\ob=0.045$, Hubble constant $h=0.7$ in units of $100\,\mathrm{km\,s^{-1}\,Mpc^{-1}}$, spectral index of the primordial power spectrum $n_s=0.96$, normalization of the present-day power spectrum $\sigma_8=0.8$, and $\fnl=0$.

With these definitions in hand, we can proceed to perform a Fisher matrix analysis \citep{Fisher:1935,Tegmark:1996bz}. Thus, the marginal error $\sigma_\fnl$ obeys
$$
\frac{1}{\sigma_\fnl}=\sqrt{\sum_{\ell=\ell_\mathrm{min}}^{\ell_\mathrm{max}}(2\ell+1)f_\mathrm{sky}\frac{\partial[C^\mathrm{HI}_\ell]_{ij}}{\partial\fnl}\left[\Gamma^\mathrm{HI}_\ell\right]^{-1}_{ij,mn}\frac{\partial[C^\mathrm{HI}_\ell]_{mn}}{\partial\fnl}},
$$
where $[C^\mathrm{HI}_\ell]_{ij}$ is a shorthand notation for $C^\mathrm{HI}_\ell(\nu_i,\nu_j)$, and $[\Gamma^\mathrm{HI}_\ell]_{ij,mn}=[C^\mathrm{HI}_\ell+\mathcal N^\mathrm{HI}_\ell]_{im}[C^\mathrm{HI}_\ell+\mathcal N^\mathrm{HI}_\ell]_{jn}+[C^\mathrm{HI}_\ell+\mathcal N^\mathrm{HI}_\ell]_{in}[C^\mathrm{HI}_\ell+\mathcal N^\mathrm{HI}_\ell]_{jm}$ is the covariance of the signal, given $\mathcal N^\mathrm{HI}_\ell$ the angular power spectrum of experimental noise (shot noise is assumed to be negligible). In this analysis we are focusing on $\fnl$ as the single parameter, a valid approximation on ultralarge scales.

\begin{figure}
\centering
\includegraphics[width=0.5\textwidth]{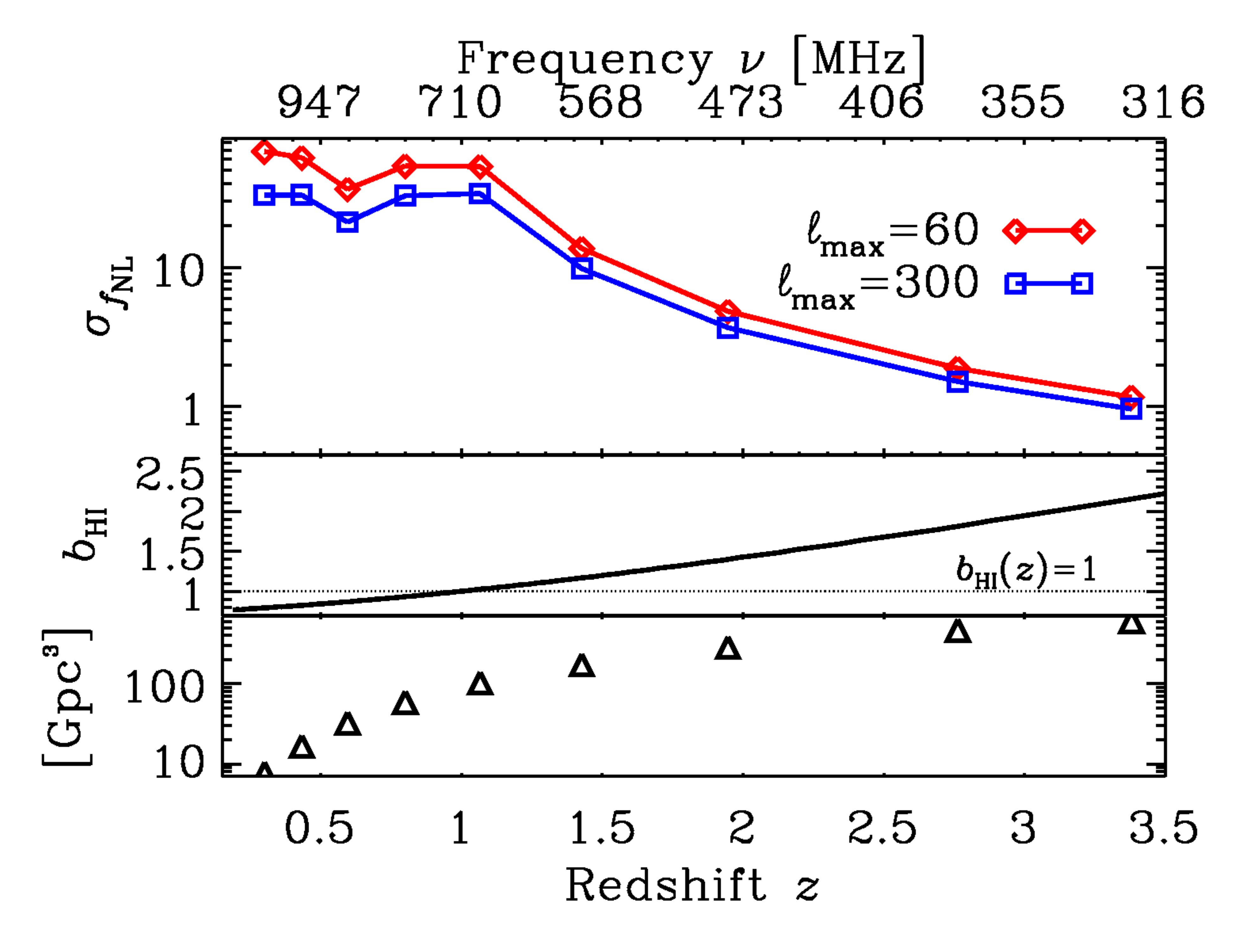}
\vskip -0.2in
\caption{Top panel: Forecasted $68.3\%$ error on $\fnl$ as a function of the mean redshift (or frequency) with a frequency interval of $\Delta\nu=200\,\mathrm{MHz}$. Middle panel: Redshift dependence of the bias. Bottom panel: Effective volume of the survey.}\label{fig:sfNL_z}
\vskip -0.2in
\end{figure}
Again, let us first focus on a zero-noise experiment. In the top panel of Fig.~\ref{fig:sfNL_z}, we show how $\sigma_\fnl$ changes with frequency range, for $\ell_\mathrm{max}=60$ (red diamonds) and $300$ (blue squares). Each point in the panel refers to the central redshift of a $200\,\mathrm{MHz}$ band with bin width $\Delta\nu=10\,\mathrm{MHz}$. As we saw before, it is essential for the bias to move away from $1$ (usually at $z \sim 1$) in order to obtain strong constraints. Moreover, it is the volume of the survey that determines the ability to probe below $\fnl$ of 10: for such a method to be successful, we need a deep survey with a large bandwidth accessing frequencies of $400\,\mathrm{MHz}$ and below. Crucially, given our fundamental ignorance about the redshift evolution of the bias, we need to span a wide range of redshifts to capture $b_\mathrm{HI}(k,z)\neq 1$.

If we are to access the cosmic variance regime, we need to explore different experimental strategies. We focus on a bandwidth $\mathrm{BW}\in[250,\,1000]\,\mathrm{MHz}$, which corresponds to $0.5\lesssim z\lesssim4.5$, subdivided into $75$ frequency bins of width $\Delta\nu=10\,\mathrm{MHz}$. This implies a $75\times75$ tomographic $[C^\mathrm{HI}_\ell]_{ij}$ matrix, that we approximate by considering block-diagonal $20\times20$ tomographic submatrices and correct for the overlap. For a survey using dishes, the expected noise can be expressed via $\mathcal N^\mathrm{HI}_\ell=T_\mathrm{sys}^2S_\mathrm{area}/(N_dt_\mathrm{TOT}\Delta\nu)$, with $T_\mathrm{sys}$ the system temperature, $S_\mathrm{area}$ the total surveyed area, $N_d$ the number of dishes, and $t_\mathrm{TOT}$ the total observation time. For interferometers we will assume that it will not be possible to mosaic. Thus, we fix the largest scale probed as the one probed by one single pointing which is set by the field of view, $\fov$. Hence, $\fov=4\pi f_\mathrm{sky}$, and we have that $\mathcal N^\mathrm{HI}_\ell=T_\mathrm{sys}^2(2\pi)^3/[f^2(\ell)\ell_\mathrm{max}^2t_\mathrm{obs}\Delta\nu]$, with $f(\ell)$ the so-called filling factor, which we fix to unity (e.g. we take a dense array). If we perform several pointing, $N_p$, such that $S_\mathrm{area}=N_p\fov$, we can replace $t_\mathrm{obs}$ with $t_\mathrm{obs}/N_p$ and divide the total $C^\mathrm{HI}_\ell+\mathcal N^\mathrm{HI}_\ell$ by $\sqrt{N_p}$. In both experimental scenarios, we adopt $T_\mathrm{sys}=[30+60\times(300\,\mathrm{MHz}/\nu)^{2.55}]\,\mathrm K$, which takes the galactic synchrotron contribution at low frequencies into account. 

While foreground cleaning will remove some information, in these experiments we have access to a very large bandwidth, Hence, we could perform the cleaning on scales which are much larger than the frequency ``chunks'' used for the cosmological analysis. For instance, in the case of reionization, Chapman \textit{et al.} \citep{Chapman:2012yj} have show that foreground cleaning will have impact {\it only} on the scales related to the size of the bandwidth used for the foreground removal---for the higher frequency range of interest to us  we expect less of an effect from the foreground removal given that the amplitude of galactic synchrotron emission will be smaller.

\begin{figure}
\vskip -0.15in
\centering
\includegraphics[width=0.5\textwidth]{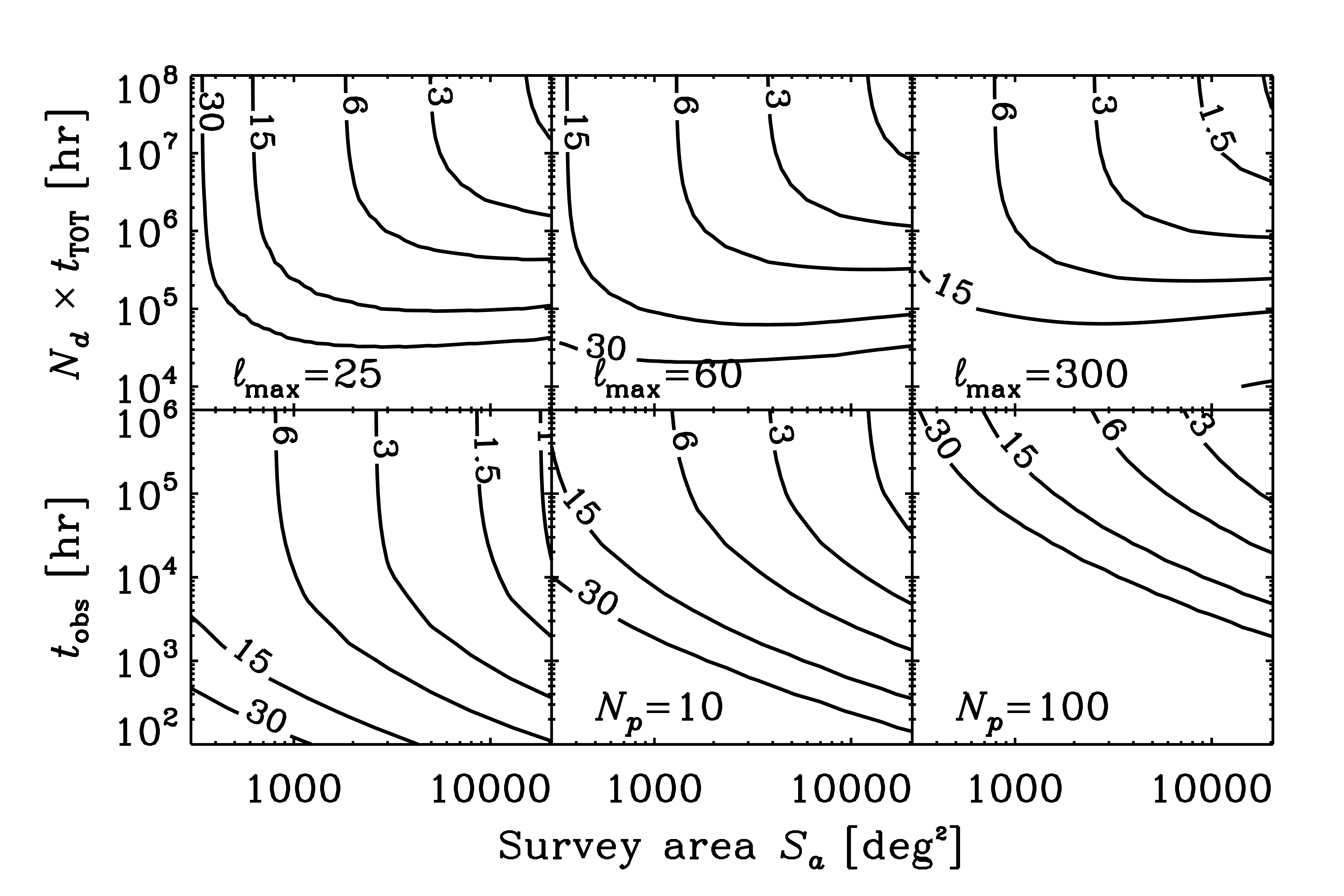}
\vskip -0.15in
\caption{Forecasted $68.3\%$ error contours on $\fnl$ as a function of surveyed area and total observation time, for a dish survey with $N_d$ dishes (upper panels) and an interferometer making $N_p$ pointings (lower panels).}\label{fig:exp}
\vskip -0.2in
\end{figure}

Fig.~\ref{fig:exp} depicts these results---we plot $\sigma_\fnl$ contours in the plane of the surveyed area and total observation time. Abscissas roughly cover from a $15\times15\,\mathrm{deg}^2$ survey to half-sky. The three top panels stand for the dish survey case, where the $y$-axis actually shows $t_\mathrm{TOT}$ multiplied by the number of dishes $N_d$. We show three maximum angular modes, namely, $\ell_\mathrm{max}=25$, $60$, and $300$ (corresponding to dish diameters of $5$, $15$ and $80$ m at redshift $\sim3$). Constraints should improve as we increase the surveyed area, since $\ell_\mathrm{min}$ decreases, thus accessing the scales which are the most affected by non-Gaussianity; on the other hand, since noise is proportional to $S_\mathrm{area}$, there will be an optimal value for it, above which errors will increase again (clearly visible for large $\sigma_\fnl$ contours).

For higher angular resolution, interferometers may be a better option.  In the lower panels of Fig.~\ref{fig:exp} we show $\sigma_\fnl$ for $1$, $10$ and $100$ pointings. Choosing $D_a\sim 80\,\mathrm m$ as the diameter for the array, the resolution is set at $\ell_\mathrm{max}\simeq300$. The main design parameter is the field of view, which sets $\ell_\mathrm{min}=2\pi/\sqrt{\fov}$ and is fixed by the effective size of each element, $d_\mathrm{eff}\sim \lambda/\sqrt{\fov}$. With a filling factor of 1, this is related to the number of elements, $N_e\sim D_a^2/d_\mathrm{eff}^2$. Given that the maximum angular scale is set by the $\fov$,  by adding more pointings, we simply diminish the variance $\Gamma^\mathrm{HI}_\ell$ by $N_p$, though the noise increases too, because $t_\mathrm{obs}\to t_\mathrm{obs}/N_p$.

There are several telescopes in development or deployed that should be able to probe this HI intensity signal and in turn help to constrain primordial non-Gaussianity. Note that in principle any interferometer can also be used as a dish experiment, as long as the autocorrelation data from each dish is saved. Moreover, the required survey can be done concurrently with any other large survey, which should increase the available observation time. It is unrealistic to expect more than $\sim10\,000\,\mathrm h$ of total time, requiring a minimum of $100$ elements for a dish survey to go beyond Planck constraints. Telescopes such as MeerKAT (http://www.ska.ac.za/meerkat) will be in this range, with its $64$ $13.5\,\mathrm m$ dishes. It will improve for SKA phase 1 with an extra $190$ $15\,\mathrm m$ dishes, also to be assembled at the same site. The minimum frequency for MeerKAT is $\sim580\,\mathrm{MHz}$. According to Fig.~\ref{fig:sfNL_z}, it means that a survey using MeerKAT will be limited to $\sigma_\fnl\sim10$. On the other hand, the SKA phase 1 dish array (middle frequencies) should probe down to $\sim 350\,\mathrm{MHz}$, thus allowing us to push below Planck constraints (using it just as a set of single dishes). Instruments such as APERTIF \citep{Oosterloo:2010wz} or ASKAP (http://www.atnf.csiro.au/projects/askap) will achieve large survey speeds thanks to the `phased array feed' system. Unfortunately, their minimum frequency is set at $1\,\mathrm{GHz}$ and $700\,\mathrm{MHz}$, respectively. This will render them unusable for probing HI at high $z$'s. Contrarily, GMRT (http://www.ncra.tifr.res.in/ncra/gmrt) can probe frequencies between $50$ and $1420\,\mathrm{MHz}$, but with only $30$ dishes and a system temperature above $100\,\mathrm K$, it will make it hard to go into the cosmic variance dominated r\'egime. In terms of designing a new system from scratch, something like $10\,000$ small dishes between $2$ and $4\,\mathrm m$ diameter, working at $\sim400\,\mathrm{MHz}$, would be a good (and cheap) possibility to target the $\fnl\sim1$ region (note that we are not requiring cross correlations between the dishes, thus making the system much simpler). This would provide an efficient experiment to probe the dark matter power spectrum out to $z\sim4$ on ultralarge scales.

For interferometric surveys, none of the planned telescopes above are compact enough to deliver the required sensitivity on large scales. This means we would need in principle to wait for SKA phase 2, with the proposed ``aperture array'' system working below $1\,\mathrm{GHz}$. Although the design is not set yet, it should be possible to achieve $\fov\sim 1000\,\mathrm{deg}^2$, thus reaching the $\sigma_\fnl\lesssim1$ limit. However, SKA phase 2 is designed to achieve much higher angular resolution than what is needed for our purposes, and a smaller array with 80 m or less in diameter would be an interesting, near term, alternative, capable of reaching $\sigma_\fnl\sim1$.

While we have outlined a simple forecasting procedure, there are two effects which must be taken into consideration if we were to lay out the specific experimental design and survey strategy. The signature we seek kicks in on scales where general relativistic and gauge effects become non-negligible. Furthermore, we have assumed that we have efficiently subtracted the foreground (i.e., our Galaxy) from the data set. We acknowledge this is an open question but, in principle, should be achievable---the Galaxy contributes a slowly varying frequency dependent signal along each line of sight, which can be accurately removed with enough radial bins to map out the small-scale structure. Indeed, foreground cleaning should be done using the largest available bandwidths ($\sim 1\,\mathrm{GHz}$) while our signal analysis will be done using smaller bandwidths. We see in Fig.~\ref{fig:sfNL_z} that $\sim200\,\mathrm{MHz}$ can be enough as long as we are probing high redshifts. Nevertheless, a detailed and realistic analysis 
of how the foreground subtraction will affect our forecast must  be undertaken. 

We have shown the strength and weaknesses of HI surveys to efficiently constrain $\fnl$. More generally, our analysis gives us an idea of how effective  HI surveys are at probing the ultralarge-scales of the cosmos and in doing so, telling us more about exotic aspects of our current cosmological models.

\textit{Acknowledgments.---}We thank R. Battye, T. Louis and R. Maartens for useful discussions. SC, MGS and LF acknowledge support from FCT-Portugal Project No. PTDC/FIS/100170/2008. SC is funded by FCT-Portugal under Post-Doctoral Grant No. SFRH/BPD/80274/2011. PGF acknowledges support from STFC, BIPAC, Leverhulme Trust and the Oxford Martin School.

\bibliographystyle{apsrev4-1}
\vskip -0.28in
\bibliography{/home/stefano/Documents/LaTeX/Bibliography}
\end{document}